\documentclass[10pt,onecolumn, a4paper]{article}
\usepackage{lmodern}
\usepackage{amssymb,amsmath}
\usepackage{ifxetex,ifluatex}
\usepackage{fixltx2e} 
\ifnum 0\ifxetex 1\fi\ifluatex 1\fi=0 
  \usepackage[T1]{fontenc}
  \usepackage[utf8]{inputenc}
\else 
  \ifxetex
    \usepackage{mathspec}
  \else
    \usepackage{fontspec}
  \fi
  \defaultfontfeatures{Ligatures=TeX,Scale=MatchLowercase}
\fi
\IfFileExists{upquote.sty}{\usepackage{upquote}}{}
\IfFileExists{microtype.sty}{%
\usepackage{microtype}
\UseMicrotypeSet[protrusion]{basicmath} 
}{}
\usepackage[margin=1in]{geometry}
\usepackage{hyperref}
\hypersetup{unicode=true,
            pdftitle={Climbing Escher's stairs: a way to approximate stability landscapes in multidimensional systems},
            pdfkeywords={quasipotential, visualization},
            pdfborder={0 0 0},
            breaklinks=true}
\usepackage{longtable,booktabs}
\usepackage{graphicx,grffile}
\makeatletter
\def\maxwidth{\ifdim\Gin@nat@width>\linewidth\linewidth\else\Gin@nat@width\fi}
\def\maxheight{\ifdim\Gin@nat@height>\textheight\textheight\else\Gin@nat@height\fi}
\makeatother
\setkeys{Gin}{width=\maxwidth,height=\maxheight,keepaspectratio}
\IfFileExists{parskip.sty}{%
\usepackage{parskip}
}{
\setlength{\parindent}{0pt}
\setlength{\parskip}{6pt plus 2pt minus 1pt}
}
\providecommand{\tightlist}{%
  \setlength{\itemsep}{0pt}\setlength{\parskip}{0pt}}
\ifx\paragraph\undefined\else
\let\oldparagraph\paragraph
\renewcommand{\paragraph}[1]{\oldparagraph{#1}\mbox{}}
\fi
\ifx\subparagraph\undefined\else
\let\oldsubparagraph\subparagraph
\renewcommand{\subparagraph}[1]{\oldsubparagraph{#1}\mbox{}}
\fi

\let\rmarkdownfootnote\footnote%
\def\footnote{\protect\rmarkdownfootnote}

\usepackage{titling}

  \title{Climbing Escher's stairs: a way to approximate stability landscapes in
multidimensional systems}
    \pretitle{\vspace{\droptitle}\centering\huge}
  \posttitle{\par}
    \author{}
    \preauthor{}\postauthor{}
      \predate{\centering\large\emph}
  \postdate{\par}
    \date{2019-04-18}

\usepackage{tikz}
\usepackage{pgfplots}
\usepackage{float}
\usetikzlibrary{shapes,arrows}
\usepackage{authblk}
\author[1]{Pablo Rodríguez-Sánchez \thanks{\texttt{pablo.rodriguezsanchez@wur.nl}}}
\author[1]{Egbert H. van Nes \thanks{\texttt{egbert.vannes@wur.nl}}}
\author[1]{Marten Scheffer \thanks{\texttt{marten.scheffer@wur.nl}}}
\affil[1]{Department of Aquatic Ecology, Wageningen University, The Netherlands}

\begin{document}
\maketitle
\begin{abstract}
Stability landscapes are useful for understanding the properties of
dynamical systems. These landscapes can be calculated from the system's
dynamical equations using the physical concept of scalar potential.
Unfortunately, for most biological systems with two or more state
variables such potentials do not exist. Here we use an analogy with art
to provide an accessible explanation of why this happens. Additionally,
we introduce a numerical method for decomposing differential equations
into two terms: the gradient term that has an associated potential, and
the non-gradient term that lacks it. In regions of the state space where
the magnitude of the non-gradient term is small compared to the gradient
part, we use the gradient term to approximate the potential as
quasi-potential. The non-gradient to gradient ratio can be used to
estimate the local error introduced by our approximation. Both the
algorithm and a ready-to-use implementation in the form of an R package
are provided.
\end{abstract}

\section{Introduction}\label{introduction}

With knowledge becoming progressively more interdisciplinary, the
relevance of science communication is rapidly increasing. Mathematical
concepts are among the hardest topics to communicate to non-expert
audiences, policy makers, and also to scientists with little
mathematical background. Visual methods are known to be successful ways
of explaining mathematical concepts and results to non-specialists.

\begin{figure}[H]

{\centering \includegraphics[width=200px]{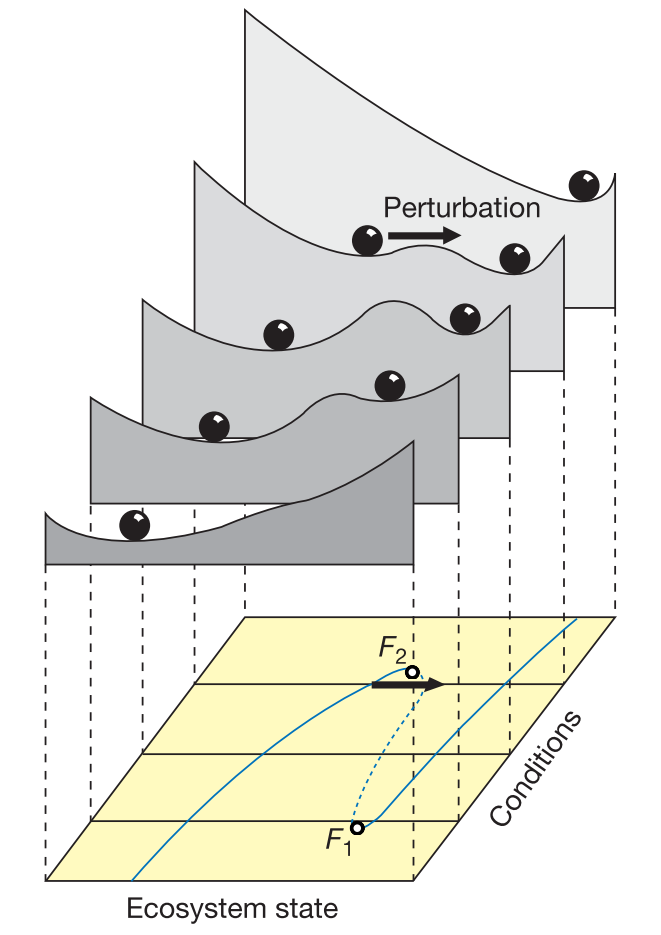} 

}

\caption{Example of a set of 4 stability landscapes (reproduced from
Scheffer et al. (2001)) used to illustrate bistability in ecosystems
(e.g: forest/desert, eutrophicated lake/clear water, etc.). The upper
side of the figure shows the stability landscape of a one-dimensional
system for 4 different values of a control parameter. The lower side
shows the bifurcation diagram. This diagram proved to be a successful
tool for explaining advanced concepts in dynamical systems theory such
as bistability and fold bifurcations to scientific communities as
diverse as ecologists, mathematicians and environmental scientists.}\label{fig:aew}
\end{figure}

One particularly successful visualization method is that of the
stability landscape, also known as the rolling marble or ball-in-a-cup
diagram Edelstein-Keshet (1988), Strogatz (1994), Beisner, Haydon, and
Cuddington (2003), Pawlowski (2006), which origin can be traced back to
the introduction of the scalar potential in physics by Lagrange in the
18th century Lagrange (1777). In stability landscapes (e.g.: figure
\ref{fig:aew}) the horizontal position of the marble represents the
state of the system at a given time. With this picture in mind, the
shape of the surface represents the underlying dynamical rules, where
the slope is proportional to the speed of the movement. The peaks on the
undulated surface represent unstable equilibrium states and the wells
represent stable equilibria. Different basins of attraction are thus
separated by peaks in the surface. Stability landscapes have proven to
be a successful tool to explain advanced concepts related with the
stability of dynamical systems in an intuitive way. Some examples of
those advanced concepts are multistability, basin of attraction,
bifurcation points and hysteresis (see Scheffer et al. (2001), Beisner,
Haydon, and Cuddington (2003) and figure \ref{fig:aew}).

The main reason for the success of this picture arises from the fact
that stability landscapes are built as an analogy with our most familiar
dynamical system: movement of a marble. Particularly, the movement of a
marble along a curved landscape under the influence of its own weight.
The stability landscape corresponds then with the physical concept of
potential energy Strogatz (1994). This explains why our intuition, based
in what we know about movement in our everyday life, works so well
reading this diagrams. It is important to stress the fact that under
this picture there's not such a thing as inertia Pawlowski (2006). The
accurate analogy is that of a marble rolling in a surface while
submerged inside a very viscous fluid Strogatz (1994).

\begin{figure}[H]

{\centering \includegraphics[width=200px]{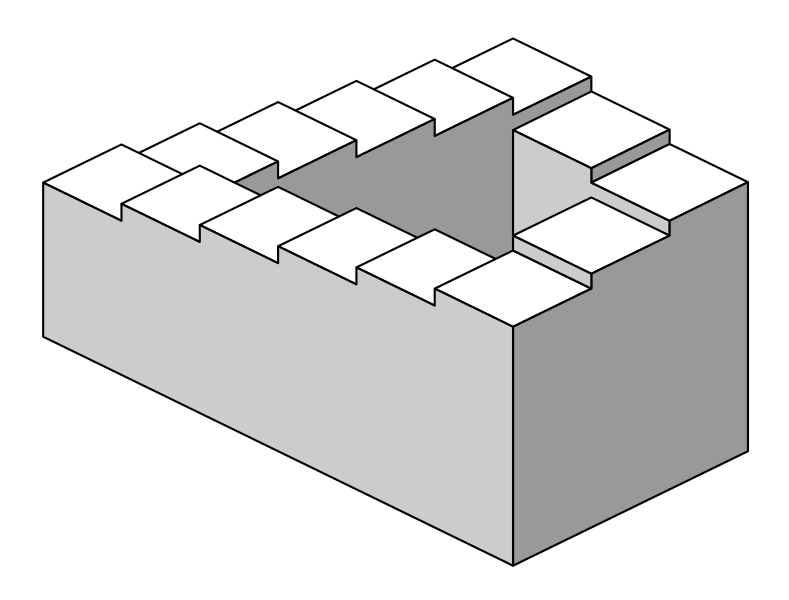} 

}

\caption{The Penrose stair L. S. Penrose and Penrose (1958) is a
classical example of an impossible object. In such a surface, it is
possible to walk in a closed loop while permanently going downhill. The
scalar potential of a system with a cyclic attractor, if existed, should
have the same impossible geometry. This object was popularized by the
Dutch artist M.C. Escher (for two beautiful examples, see Escher (1960)
and Escher (1961)).}\label{fig:penrose}
\end{figure}

Like with any other analogy, it is important to be aware of its
limitations. The one we address here is the fact that, for models with
more than one state variable, such a potential doesn't exist in general.
To get an intuitive feeling of why this is true, picture a model with a
stable cyclic attractor. As the slope of the potential should reflect
the speed of change, we would need a potential landscape where our
marble can roll in a closed loop while always going downhill. Such a
surface is a classical example of an impossible object (see figure
\ref{fig:penrose} and L. S. Penrose and Penrose (1958) for details).

As this is a centuries-old problem (see for instance Helmholtz (1858)),
it is perhaps of not surprise that several methods have been proposed to
approximate stability landscapes for general, high-dimensional systems.
For a comprehensive mathematical review, we refer to Zhou et al. (2012).
Here, different criteria are used to decompose analytically the dynamics
in a scalar potential and an extra term. Some interesting alternatives
are presented in Pawlowski (2006), like potentials for second-order
systems or the use of Lyapunov functions as stability landscapes. For
stochastic systems, the Freidlin-Wentzell potential Freidlin and
Wentzell (2012) has been proposed as a very strong candidate due to its
links with transition rates Zhou et al. (2012), Nolting and Abbott
(2015).

In the present work we review the state of the art, and we introduce a
simple new method to deal with the fundamental problem of approximating
stability landscapes for high dimensional systems. Specificially, we
introduce an algorithm to decompose differential equations as the sum of
a gradient and a non-gradient part. Each part can be used, respectively,
to compute an associated potential and to measure the local error
introduced by our picture. In order to reach those interested readers
with little background in mathematics, we limited our mathematical
weaponry. Knowledge of basic linear algebra and basic vector calculus
will suffice to understand the paper to its last detail. Additionally,
we provide a ready to use R package that implements the algorithm this
paper describes.

\subsection{Mathematical background}\label{mathematical-background}

Consider a coupled differential equation with two state variables \(x\)
and \(y\) (see online appendix \ref{sec:generalization} for a
generalization in more dimensions). The dynamics of such system can be
described by a stability landscape if a single two-dimensional function
\(V(x,y)\) exists whose slope is proportional to the change in time of
both states (equation \eqref{eq:2D}).

\begin{equation}
  \begin{cases}
    \frac{dx}{dt} = f(x,y) = - \frac{\partial V(x,y)}{\partial x} \\
    \frac{dy}{dt} = g(x,y) = - \frac{\partial V(x,y)}{\partial y}
  \end{cases}
  \label{eq:2D}
\end{equation}

It can be shown that such potential \(V(x,y)\) only exists if the
crossed derivatives of functions \(f(x,y)\) and \(g(x,y)\) are equal for
all \(x\) and \(y\) (equation \eqref{eq:2Dcond}). Systems satisfying
equation \eqref{eq:2Dcond} are known as conservative, irrotational or
gradient fields (cf.~section 8.3 of Marsden and Tromba (2003)).

\begin{equation}
\frac{\partial f}{\partial y} = \frac{\partial g}{\partial x}
\label{eq:2Dcond}
\end{equation}

If condition \eqref{eq:2Dcond} holds, we can use a line integral (Marsden
and Tromba (2003), section 7.2) to invert \eqref{eq:2D} and calculate
\(V(x,y)\) using the functions \(f(x,y)\) and \(g(x,y)\) as an input. An
example of this inversion is equation \eqref{eq:2Dint}, where we have
chosen an integration path composed of a horizontal and a vertical line.

\begin{equation}
V(x,y) = V(x_0, y_0) - \int^{x}_{x_0} f(\xi, y_0) d\xi - \int^{y}_{y_0} g(x, \eta) d\eta
\label{eq:2Dint}
\end{equation}

The attentive reader may have raised her or his eyebrow after reading
the word \emph{chosen} applied to an algorithm. In fact, we can
introduce this arbitrary choice without affecting the final result. The
condition for potentials to exist (equation \eqref{eq:2Dcond}) implies
that any line integral between two points in this vector field should be
independent of the path (cf.~section 7.2 of Marsden and Tromba (2003)).
Going back to our rolling marble analogy, we can gain some intuition
about why this is true: in a landscape the difference in potential
energy between two points is proportional to the difference in height,
and thus stay the same for any path. If the condition \eqref{eq:2Dcond})
is not fulfilled, the potential calculated with \eqref{eq:2Dint}) will
depend on the chosen integration path. As this is an arbitrary choice,
the computed potential will be an artifact with no natural meaning.

For the sake of a compact and easy to generalize notation, in the rest
of this paper we will arrange the equations of our system as a column
vector (equation \eqref{eq:2Dfieldnotation}).

\begin{equation}
  \begin{bmatrix}
           f(x,y) \\
           g(x,y)
  \end{bmatrix}
=
\vec f (\vec x)
\label{eq:2Dfieldnotation}
\end{equation}

\section{Methods}\label{methods}

The method for deriving a potential we propose is based on the
decomposition of a vector field in a conservative or gradient part and a
non-gradient part (see equation \eqref{eq:FieldDecomposition}).

\begin{equation}
\vec f (\vec x) = \vec f_g (\vec x) + \vec f_{ng} (\vec x)
\label{eq:FieldDecomposition}
\end{equation}

The gradient term (\(\vec f_g (\vec x)\)) captures the part of the
system that can be associated to a potential function, while the
non-gradient term (\(\vec f_{ng} (\vec x)\)) represents the deviation
from this ideal case. We'll use \(\vec f_g (\vec x)\) to compute an
approximate or quasi-potential. The absolute error of this approach is
given as the euclidean size of the non-gradient term
\(\mid \vec f_{ng} (\vec x) \mid\). In regions where the gradient term
is stronger than the non-gradient term, the condition \eqref{eq:2Dcond}
will be approximately fulfilled, and thus the calculated quasi-potential
will represent an acceptable approximation of the underlying dynamics.
Otherwise, the non-gradient term is too dominant to approximate a
potential landscape.

In order to achieve a decomposition like \eqref{eq:FieldDecomposition}, we
begin by linearizing our model equations. Any sufficiently smooth and
continuous vector field \(\vec f(\vec x)\) can be approximated around a
point \(\vec{x_0}\) using equation \eqref{eq:TaylorExp}, where
\(J(\vec{x_0})\) is the Jacobian matrix evaluated at the point
\(\vec{x_0}\) and \(\Delta \vec x\) is defined as the distance to this
point, that is, \(\Delta \vec x = \vec x - \vec{x_0}\), written as a
column vector.

\begin{equation}
\vec f(\vec x) \approx \vec f(\vec{x_0}) + J(\vec{x_0}) \Delta \vec x
\label{eq:TaylorExp}
\end{equation}

As usual in linearization, we have neglected the terms of order \(2\)
and higher in equation \eqref{eq:TaylorExp}. This approximation is valid
for \(\vec{x}\) close to \(\vec{x_0}\). For the system to be gradient
(equation \eqref{eq:2Dcond}) its Jacobian has to be symmetric (see
equation \eqref{eq:jaccond}, where \(T\) represents transposition).

\begin{equation}
J = J^T
\label{eq:jaccond}
\end{equation}

We know from basic linear algebra that any square matrix \(M\) can be
uniquely decomposed as the sum of a skew and a symmetric matrix (see
equation \eqref{eq:SkewSymDec}).

\begin{equation}
  \begin{cases}
    M_{symm} = \frac{1}{2} \left( M + M^T \right) \\
    M_{skew} = \frac{1}{2} \left( M - M^T \right)
  \end{cases}
\label{eq:SkewSymDec}
\end{equation}

Using the skew-symmetric decomposition described in equation
\eqref{eq:SkewSymDec}, we can rewrite \eqref{eq:TaylorExp} as:

\begin{equation}
\vec f(\vec x) \approx \vec f(\vec{x_0}) + J_{symm}(\vec{x_0}) \Delta \vec x + J_{skew}(\vec{x_0})  \Delta \vec x
\label{eq:TaylorExpDecomposed}
\end{equation}

Equation \eqref{eq:TaylorExpDecomposed} represents a natural, well-defined
and operational way of writing our vector field \(\vec f(\vec x)\)
decomposed as in equation \eqref{eq:FieldDecomposition}, that is, as the
sum of a gradient and a non-gradient term (see \eqref{eq:decomposition}).

\begin{equation}
  \begin{cases}
    \vec f_g (\vec x) \approx \vec f(\vec{x_0}) + J_{symm}(\vec{x_0})  \Delta \vec x \\
    \vec f_{ng} (\vec x) \approx J_{skew}(\vec{x_0})  \Delta \vec x
  \end{cases}
\label{eq:decomposition}
\end{equation}

The gradient term \(f_g (\vec x)\) can thus be associated to a potential
\(V(\vec x)\) that can be computed analytically for this linearized
model using a line integral (see equation \eqref{eq:2Dint} for the two
dimensional case, or \eqref{eq:anyDint} in the online appendix for the
general one). The result of this integration yields an analytical
expression for the potential difference between the reference point
\(\vec{x_0}\) and another point
\(\vec{x_1} \equiv \vec{x_0} + \Delta \vec{x}\) separated by a distance
\(\Delta \vec x\) (see equation \eqref{eq:Potential}).

\begin{equation}
\Delta V(\vec{x_1}, \vec{x_0}) \equiv V(\vec{x_1}) - V(\vec{x_0}) \approx - \vec f(\vec x_0) \cdot \Delta \vec x - \frac{1}{2} \Delta \vec x^T  J_{symm} (\vec{x_0})  \Delta \vec x
\label{eq:Potential}
\end{equation}

Provided we know the value of the potential at one point
\(\vec x_{previous}\), equation \eqref{eq:Potential} allows us to estimate
the potential at a different point \(\vec x_{next}\) (cf.: equation
\eqref{eq:Iterator}).

\begin{equation}
V(\vec x_{next}) \approx V(\vec x_{previous}) + \Delta V(\vec x_{next}, \vec x_{previous})
\label{eq:Iterator}
\end{equation}

Equation \eqref{eq:Iterator} can be applied sequentially over a grid of
points to calculate the approximate potential on each of them. In two
dimensions, the resulting potential is given by the closed formula
\eqref{eq:NumericalRecipe}. The cases with 3 and more dimensions can be
generalized straightforwardly. For a step by step example, see section
\ref{sec:2d-example} in the online appendix. For a flowchart overview of
the algorithm, please refer to figure \ref{fig:detail}.

\begin{equation}
V(x_i, y_j) = V(x_0, y_0) + \sum_{k = 1}^i \Delta V(x_k, y_0; x_{k-1}, y_0) + \sum_{l = 1}^j \Delta V(x_i, y_l; x_i, y_{l-1})
\label{eq:NumericalRecipe}
\end{equation}

As with any other approximation we need a way to estimate and control
its error. The stability landscape described in \eqref{eq:Potential} has
two main sources of errors:

\begin{enumerate}
\def\labelenumi{\arabic{enumi}.}
\tightlist
\item
  It has been derived from a set of linearized equations, sampled over a
  grid
\item
  It completely neglects the effects of the non-gradient part of the
  system
\end{enumerate}

The error due to linearization in equation \eqref{eq:Potential} is roughly
proportional to \(\mid \Delta \vec{x} \mid^3\), where
\(\mid \Delta \vec{x} \mid\) is the euclidean distance to the reference
point. By introducing a grid, we expect the linearization error to
decrease with the grid's step size.

The more fundamental error due to neglecting the non-gradient component
of our system is not affected by the grid's step choice. From equation
\eqref{eq:FieldDecomposition} it is apparent that we can use the euclidean
size of \(\vec f_{ng} (\vec x)\) as an approximation of the local error
introduced by our algorithm. The relative error due to this effect can
be estimated using equation \eqref{eq:error}.

\begin{equation}
  err(\vec x) \approx \frac{\mid \vec f_{ng}(\vec x) \mid}{\mid \vec f_g(\vec x) \mid + \mid \vec f_{ng}(\vec x) \mid} \approx \frac{\mid J_{skew}(\vec{x}) \mid}{\mid J_{symm}(\vec{x}) \mid + \mid J_{skew}(\vec{x}) \mid}
\label{eq:error}
\end{equation}

\subsection{Implementation}\label{implementation}

\begin{figure}[H]

{\centering \includegraphics[width=400px]{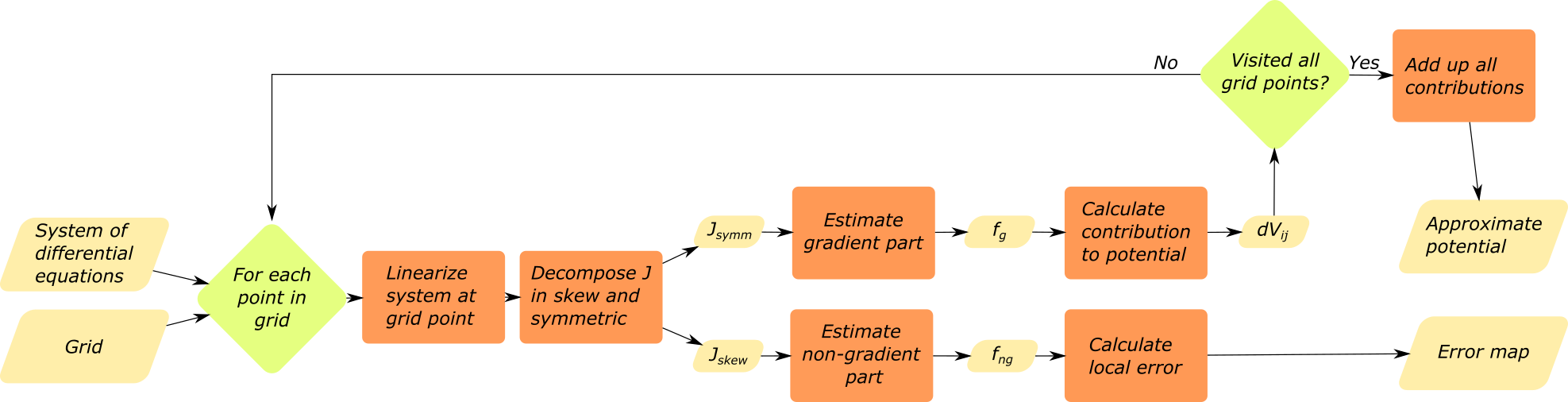} 

}

\caption{Flowchart showing the basic functioning of our
implementation of the algorithm described in this paper.}\label{fig:detail}
\end{figure}

As an application of the above-mentioned ideas, and following the spirit
of reproducible research, we developed a transparent \emph{R} package we
called \emph{rolldown} Rodríguez-Sánchez (2019). Our algorithm accepts a
set of dynamical equations and a grid of points defining our region of
interest as an input. The output is the estimated potential and the
estimated error, both of them calculated at each point of our grid (see
figure \ref{fig:detail} for details).

\section{Results}\label{results}

\subsection{Synthetic examples}\label{synthetic-examples}

\subsubsection{A four well potential}\label{a-four-well-potential}

We first tested our algorithm with a synthetic model of two uncoupled
state variables. Uncoupled systems are always gradient as all
non-diagonal values of the Jacobian are zero everywhere. We added the
interaction terms \(p_x\) and \(p_y\) to be able to make it gradually
non-gradient (see equation \eqref{eq:TestField}).

\begin{equation}
  \begin{cases}
    \frac{dx}{dt} = f(x,y) = -x (x^2 - 1) + p_x(x,y)\\
    \frac{dy}{dt} = g(x,y) = -y (y^2 - 1) + p_y(x,y)
  \end{cases}
\label{eq:TestField}
\end{equation}

\begin{figure}[H]
\includegraphics[width=400px]{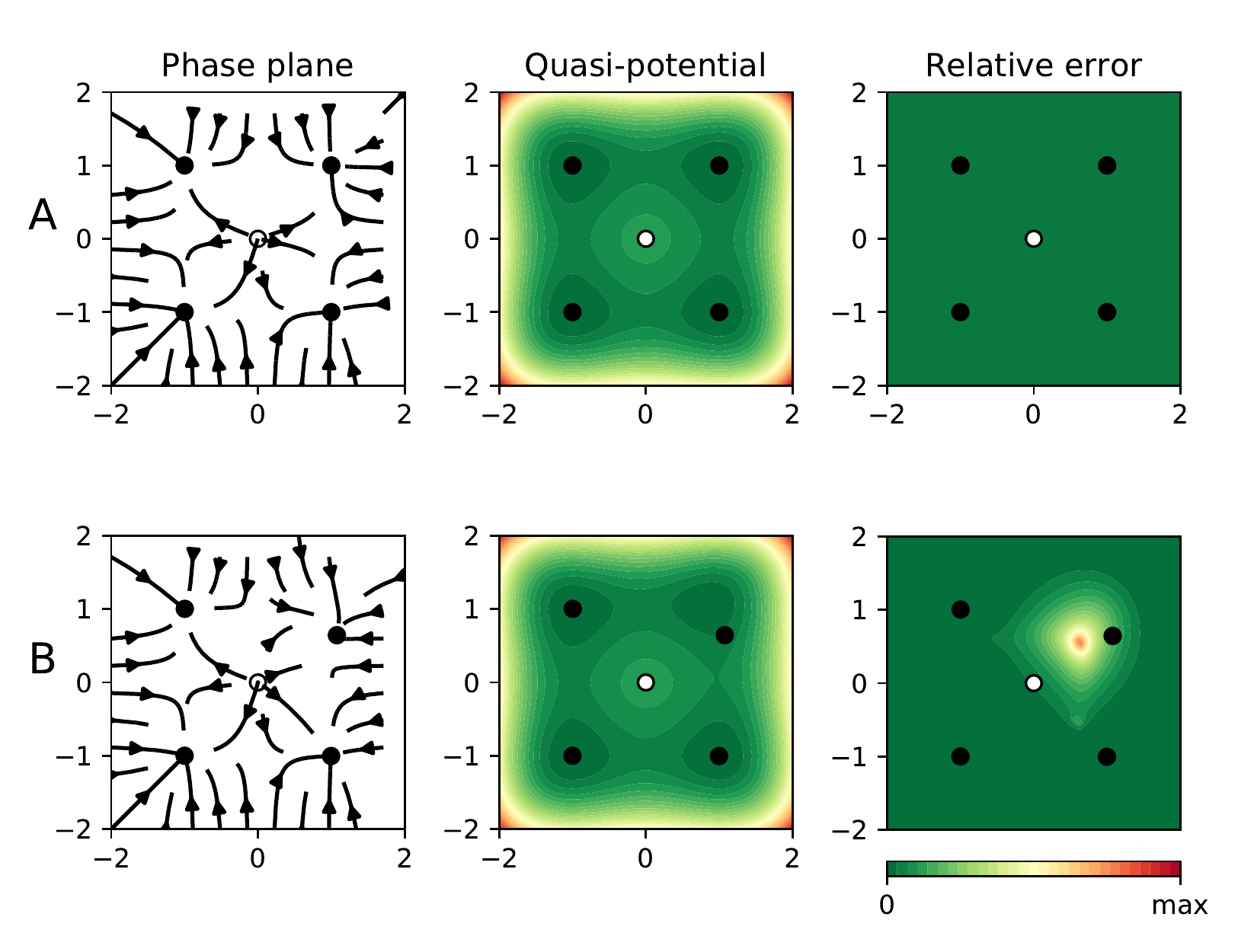} \caption{Results for two synthetic examples. In all panels
the dots represent equilibrium points (black for stable, otherwise
white). The left panel shows the phase plane containing the actual
``deterministic skeleton'' of the system. The central panel shows the
quasi-potential. The right panel shows the estimated error. \textbf{Row
A} shows the application to a gradient case (equation \eqref{eq:TestField}
with interaction terms equal to zero). As expected, the error is zero
everywhere. In \textbf{row B} our algorithm is applied to a non-gradient
case (equation \eqref{eq:TestField}, with non-zero interaction terms).}\label{fig:panel-four-py}
\end{figure}

When we choose those non-gradient interactions to be zero, the system is
purely gradient and corresponds with a four-well potential. Our
algorithm rendered it successfully and with zero error (cf.~figure
\ref{fig:panel-four-py}, row A). In order to test our algorithm, we
introduced a non-gradient interaction of the form
\(p_x(x,y) = 0.3 y \cdot m(x, y)\) and
\(p_y(x,y) = -0.4 x \cdot m(x,y)\), with
\(m(x,y) = e^{(x-1)^2 + (y-1)^2}\). \(m(x,y)\) serves as a masking
function guaranteeing that our interaction term will be negligible
everywhere but in the vicinity of \((x,y) = (1,1)\). After introducing
this non-gradient interaction a four-well potential is still
recognizable (cf.~figure \ref{fig:panel-four-py}, row B). As expected,
the error was correctly captured to be zero everywhere but in the region
around \((x,y) = (1,1)\). The error map warns us against trusting the
quasi-potential in the upper right region, and guarantees that elsewhere
it will work fine. Notice that, accordingly, the upper right stable
equilibrium falls slightly outside its corresponding well. The rest of
equilibria, to the contrary, fit correctly inside their corresponding
wells.

\subsubsection{A fully non-gradient
system}\label{a-fully-non-gradient-system}

\begin{figure}[H]

{\centering \includegraphics[width=400px]{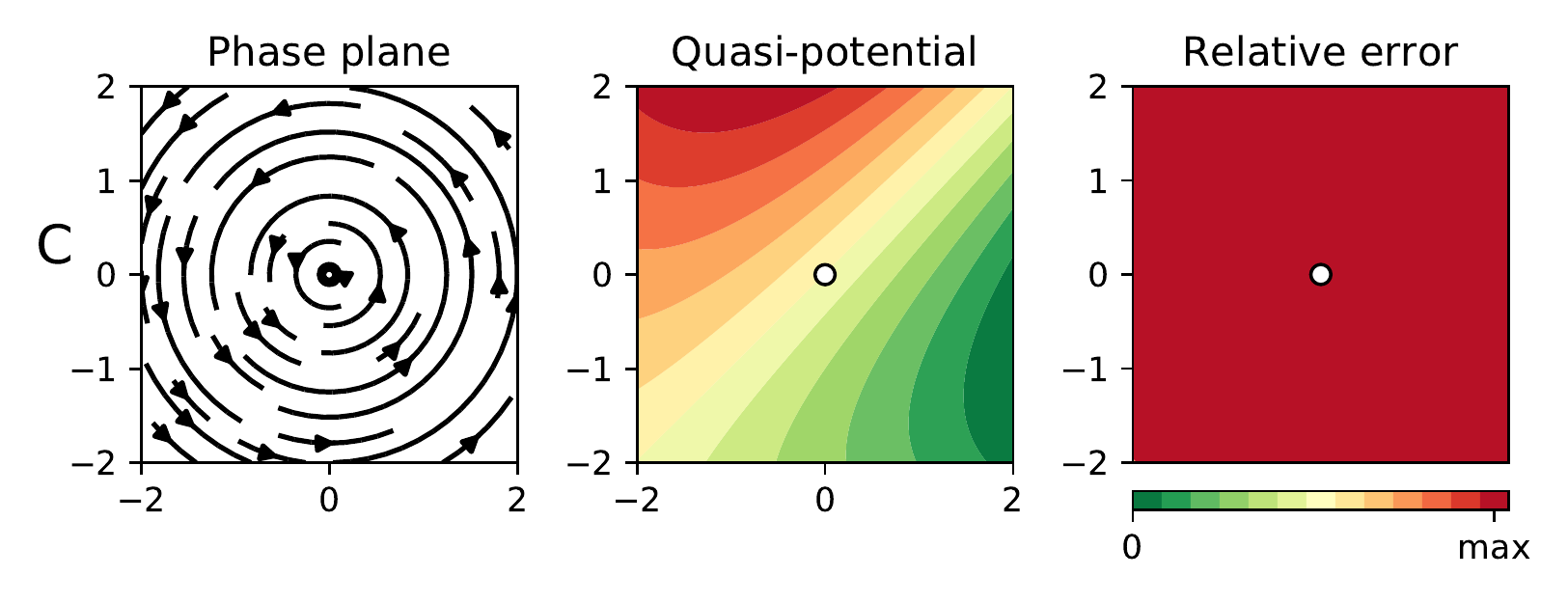} 

}

\caption{Results for a fully non-gradient system (equation
\eqref{eq:pure-curl}). In all panels the dots represent center equilibrium
points. The left panel shows the phase plane containing the actual
``deterministic skeleton'' of the system. The central panel shows the
quasi-potential. The right panel shows the estimated error.}\label{fig:panel-curl-py}
\end{figure}

In order to stress our algorithm to the maximum, we tested it in a worst
case scenario: a system with zero gradient part everywhere.
Particularly, we fed it with equation \eqref{eq:pure-curl}. All solutions
(but the unstable equilibrium point at \((0,0)\)) are cyclic (cf.
\ref{fig:panel-curl-py}, left panel)). As we discussed in the
introduction, calculating a potential for a system with cyclic
trajectories is as impossible as Escher's paintings (and for similar
reasons). This fact is captured by our algorithm, that correctly
predicts a relative error of \(1\) everywhere (see figure
\ref{fig:panel-curl-py}, central panel). In this case, the
quasi-potential (figure \ref{fig:panel-curl-py}, left panel) is not even
locally useful.

\begin{equation}
  \begin{cases}
    \frac{dx}{dt} = -y \\
    \frac{dy}{dt} = x
  \end{cases}
\label{eq:pure-curl}
\end{equation}

\subsection{Biological examples}\label{subsec:Biological-examples}

\subsubsection{A simple regulatory gene
network}\label{a-simple-regulatory-gene-network}

Waddington's epigenetic landscapes Gilbert (1991) are a particular
application of stability landscapes to gene regulatory networks
controlling cellular differentiation. When applied to this problem,
stability/epigenetic landscapes serve as a visual metaphor for the
branching pathways of cell fate determination Bhattacharya, Zhang, and
Andersen (2011).

A bistable network cell fate model can be described by a set of
equations like \eqref{eq:Waddington}. Such a system represents two genes
(\(x\) and \(y\)) that inhibit each other. This circuit works as a
toggle switch with two stable steady states, one with dominant \(x\),
the other with dominant \(y\) (see Bhattacharya, Zhang, and Andersen
(2011)).

\begin{equation}
  \begin{cases}
    \frac{dx}{dt} = b_x - r_x x + \frac{a_x}{k_x + y^n} \\
    \frac{dy}{dt} = b_y - r_y y + \frac{a_y}{k_y + x^n}
  \end{cases}
\label{eq:Waddington}
\end{equation}

Our parameter choice (\(a_x =\) 0.4802, \(a_y =\) 0.109375, \(b_x =\)
0.2, \(b_y =\) 0.3, \(k_x =\) 0.2401, \(k_y =\) 0.0625, \(r_x = r_y =\)
1 and , \(n =\) 4) corresponds with equations 6 and 7 of Bhattacharya,
Zhang, and Andersen (2011), where we modified two parameters
(\(By = 0.3\), \(foldXY = 1.75\), in their notation) in order to induce
an asymmetry in the the dynamics. Despite this system is non-gradient,
our algorithm correctly predicts the existence of two wells (see figure
\ref{fig:panel-bio-py}, row D). The relative error, despite being
distinct from zero in some regions, is not very high. This means that
our quasi-potential is a reasonable approximation of the underlying
dynamics. Indeed, the equilibria correspond to the wells (stable) and
the peak (unstable).

\subsubsection{Predator prey dynamics}\label{predator-prey-dynamics}

The Lotka-Volterra equations \eqref{eq:Lotka} are a classical
predator-prey model Volterra (1926). In this model \(x\) and \(y\)
represent, respectively, the prey and predator biomasses.

\begin{equation}
  \begin{cases}
    \frac{dx}{dt} = a x - b x y \\
    \frac{dy}{dt} = c x y - d y
  \end{cases}
\label{eq:Lotka}
\end{equation}

\begin{figure}[H]
\includegraphics[width=400px]{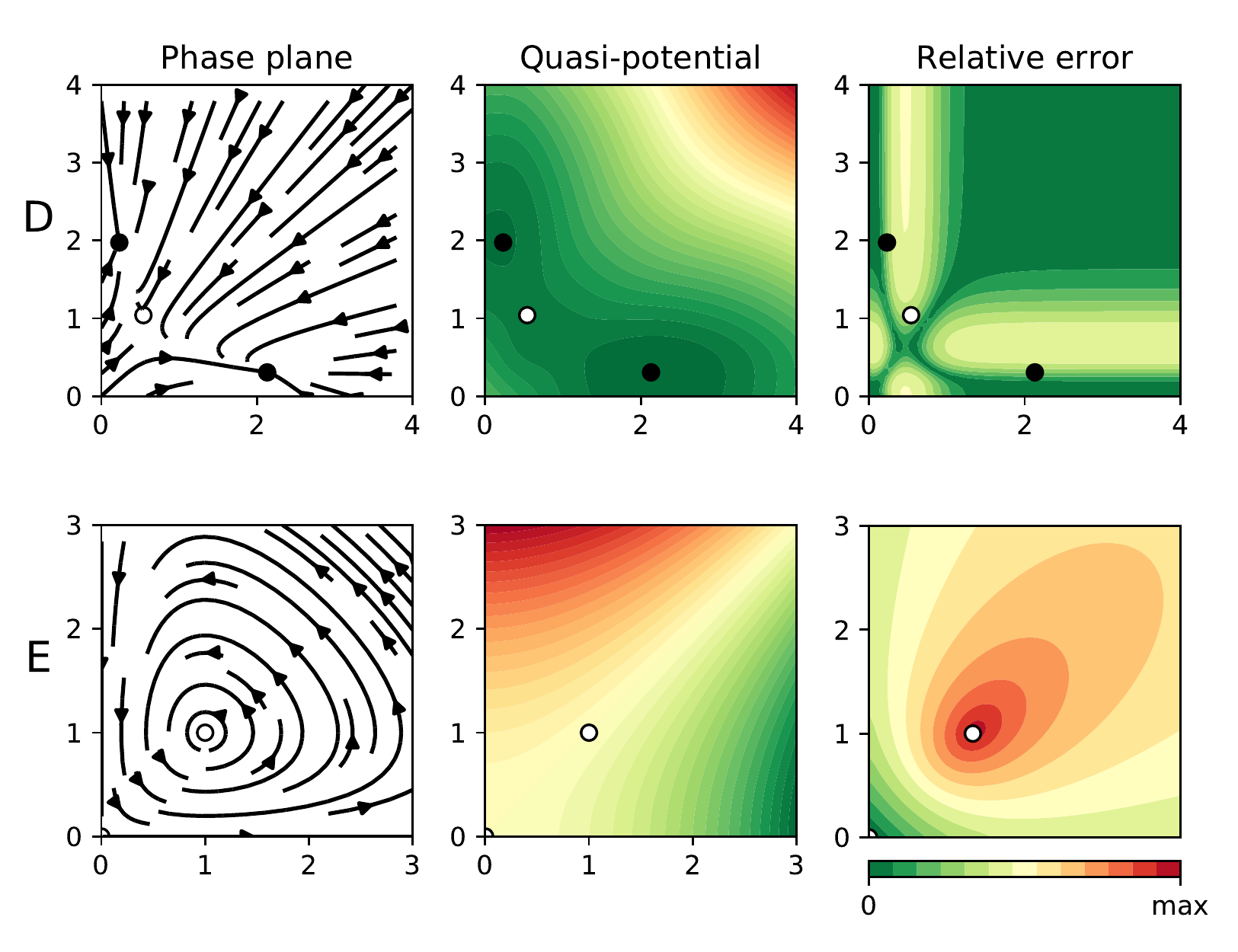} \caption{Results for two biological systems. In all panels the
dots represent equilibrium points (black for stable, otherwise white).
The left panel shows the phase plane containing the actual
``deterministic skeleton'' of the system. The central panel shows the
quasi-potential. The right panel shows the estimated error. In
\textbf{row D} we applied our algorithm to the simple gene regulatory
network described in equation \eqref{eq:Waddington}. In \textbf{row E} we
apply our algorithm to a Lotka-Volterra system (equation \eqref{eq:Lotka},
with \(a = b = c = d = 1\)).}\label{fig:panel-bio-py}
\end{figure}

This model is known to have cyclic dynamics. As we discussed in our
analogy with Escher's paintings, we cannot compute stability landscapes
in the regions of the phase plane where the dynamics are cyclic. When we
apply our method to a system like equation \eqref{eq:Lotka}, the error map
correctly captures the fact that our estimated potential is not
trustworthy (see figure \ref{fig:panel-bio-py}, row E).

\section{Discussion}\label{discussion}

The use of stability landscapes as a helping tool to understand
one-dimensional dynamical systems achieved great success, especially in
interdisciplinary research communities. A generalization of the idea of
scalar potential to two-dimensional systems seemed to be a logical next
step. Unfortunately, as we have seen, there are reasons that make two
(and higher) dimensional systems fundamentally different from the
one-dimensional case. The generalization, straightforward as it may
look, is actually impossible for most dynamical equations. A good
example of this impossibility is any system with cyclic dynamics, whose
scalar potential should be as impossible a the Penrose stairs in
Escher's paintings.

As a consequence, any attempt of computing stability landscapes for
high-dimensional systems should, necessarily, drop some of the desirable
properties of classical scalar potentials. For instance, the method
proposed by Bhattacharya Bhattacharya, Zhang, and Andersen (2011)
smartly avoids the problem of path dependence of line integrals by
integrating along trajectories, removing thus the freedom of path
choice. The price paid is that Bhattacharya's algorithm cannot guarantee
continuity along separatrices between basins of attraction for general
multistable systems.

We share the perception with other authors Pawlowski (2006) that the
concept of potential is often misunderstood in research communities with
a limited mathematical background. Analytical methods like the ones
described in Zhou et al. (2012) or Nolting and Abbott (2015) require
familiarity with advanced mathematical concepts such as partial
differential equations. The algorithm we present here is an attempt to
preserve as much as possible from the classical potential theory while
keeping the mathematical complexity as low as possible. Additionally,
our algorithm provides:

\begin{itemize}
\tightlist
\item
  Integrity. At each step the strength of the non-gradient term is
  calculated. If this term is high, it is fundamentally impossible to
  calculate a scalar potential with any method. If this term is zero,
  our solution converges to the classical potential.
\item
  Safety. The relative size of the non-gradient term can be interpreted
  as an error term, mapping which regions of our stability landscape are
  dangerous to visit.
\item
  Speed. The rendering of a printing quality surface can be performed in
  no more than a few seconds in a personal laptop.
\item
  Simplicity. The required mathematical background is covered by any
  introductory course in linear algebra and vector calculus.
\item
  Generality. The core of the algorithm is the skew-symmetric
  decomposition of the Jacobian. This operation can be easily applied to
  square matrices of any size, generalizing our algorithm for working in
  3 or more dimensions.
\item
  Usability. We provide the algorithm in the form of a ready to use,
  documented and tested \emph{R} package.
\end{itemize}

It is important to notice that, despite our algorithm provides us with a
way of knowing which regions of the phase plane can be ``safely
visited'', we cannot navigate the phase plane freely but only along
trajectories. This interplay between regions and trajectories limits the
practical applicability of our algorithm to those trajectories that
never enter regions with high error.

The concept of potential is paramount in physical sciences. The main
reason for the ubiquity of potentials in physics is that several
(idealized) physical systems are known to be governed only by gradient
terms (e.g.: movement in friction-less systems, classical gravitatory
fields, electrostatic fields, \ldots{}). As physical potentials can be
related with measurable concepts like energy, its use goes way further
than visualization. From the depth and width of a potential we can learn
about transition rates and resilience to pulse perturbations. The height
of the lowest barrier determines the minimum energy to transition to an
alternative stable state, which relates to the probability of a critical
transition in a stochastic environment Zhou et al. (2012), Nolting and
Abbott (2015), Hänggi, Talkner, and Borkovec (1990). All these results
remain true for non-physical problems that happen to be governed
exclusively by gradient dynamics, and, we claim, should remain
approximately true for problems governed by weakly non-gradient
dynamics. This is the situation our algorithm has been designed to deal
with.

Regarding visualization alone, it may be worth reconsidering why do we
prefer the idea of stability landscape over a traditional phase plane
figure, especially after pointing out all the difficulties of
calculating stability landscapes for higher-dimensional systems. It is
true that the phase plane is slightly less intuitive than the stability
landscape, but it has a very desirable property: it doesn't require the
imagination of a surrealist artist to exist.

\section{Acknowledgments}\label{acknowledgments}

This work was greatly inspired by the discussions with Cristina Sargent,
Iñaki Úcar, Enrique Benito, Sanne J.P. van den Berg, Tobias
Oertel-Jäger, Jelle Lever, and Els Weinans. This work was supported by
funding from the European Union's \emph{Horizon 2020} research and
innovation programme for the \emph{ITN CRITICS} under Grant Agreement
Number \emph{643073}.

\newpage

\section*{Online appendix}\label{online-appendix}
\addcontentsline{toc}{section}{Online appendix}

\setcounter{equation}{0} \setcounter{figure}{0} \setcounter{section}{0}
\setcounter{subsection}{0}
\renewcommand{\theequation}{A.\arabic{equation}}
\renewcommand{\thefigure}{A.\arabic{figure}}
\renewcommand{\thesubsection}{A.\arabic{subsection}}

\subsection{Gradient conditions for a system with an arbitrary number
of dimensions}\label{sec:generalization}

Dynamics in equation \eqref{eq:2D} and the condition for the crossed
derivatives \eqref{eq:2Dcond} can be straightforwardly generalized (see
equations \eqref{eq:anyD} and \eqref{eq:anyDcond} to systems with an
arbitrary number of state variables \(\vec x = (x_1, ..., x_n)\).
Particularly, if and only if our system of equations
\(\frac{dx_i}{dt} = f_i(\vec x)\) satisfies the condition:

\begin{equation}
  \frac{\partial f_i}{\partial x_j} = \frac{\partial f_j}{\partial x_i} : i \neq j
\label{eq:anyDcond}
\end{equation}

then a potential \(V(\vec x)\) exists related to the original vector
field:

\begin{equation}
  \frac{d x_i}{dt} = f_i(\vec x) = -\frac{\partial V}{\partial x_i} : i = 1..n
\label{eq:anyD}
\end{equation}

and such a potential can be computed using a line integral:

\begin{equation}
V(\vec x) = V(\vec x_0) - \int_\Gamma \sum_{i=1}^n f_i(\vec x) dx_i
\label{eq:anyDint}
\end{equation}

where the line integral in \eqref{eq:anyDint} is computed along any curve
\(\Gamma\) joining the points \(\vec x_0\) and \(\vec x\).

It is important to note that the number of equations (\(N\)) described
in condition \eqref{eq:anyDcond} grows rapidly with the dimensionality of
the system (\(D\)), following the series of triangular numbers
\(N = \frac{1}{2}(D-1)D\). Thus, the higher the dimensionality, the
harder it may get to fulfill condition \eqref{eq:anyDcond}. As a side
effect, we see that one-dimensional systems have zero conditions and
their stability landscape is thus always well-defined.

\subsection{Detailed example of application}\label{sec:2d-example}

\begin{figure}[H]

{\centering \includegraphics{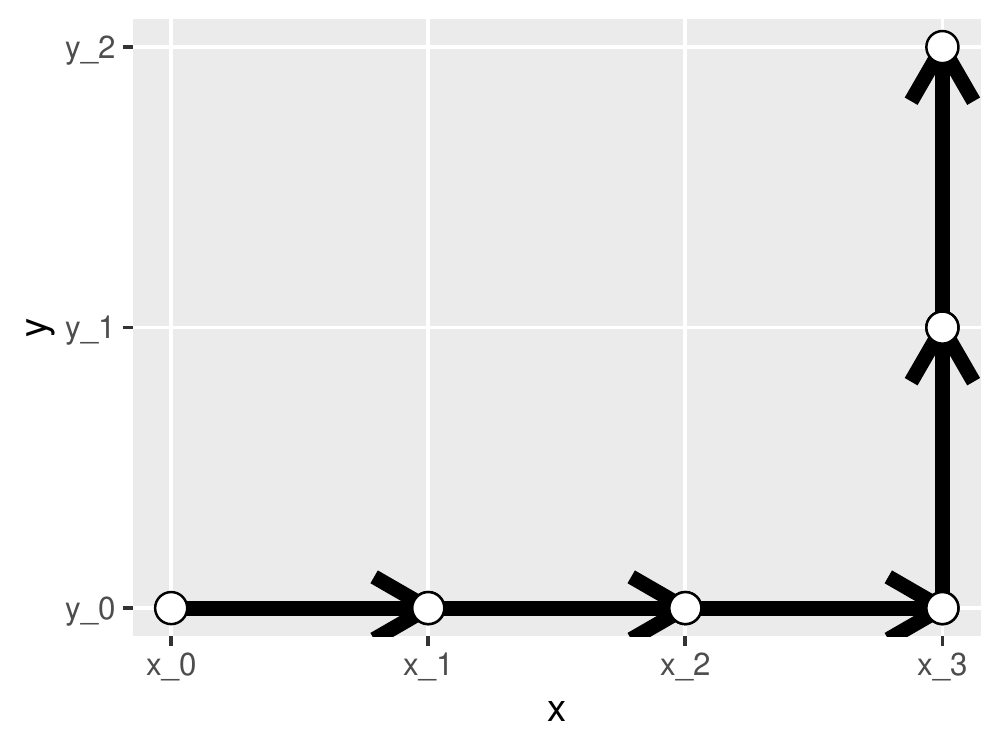} 

}

\caption{Path used to go from point \((x_0,y_0)\) to \((x_3, y_2)\).
Note that this is not the only possible path. Our algorithm converges to
the same potential regardless of the path chosen thanks to neglecting
the skew part of the Jacobian in our linearization process.}\label{fig:steps}
\end{figure}

To calculate the value of \(V\) at, for instance, the point
\((x_3, y_2)\) of a grid, we should begin by assigning \(0\) to the
potential at our arbitrary starting point (i.e.: \(V(x_0, y_0) = 0\) by
definition). Then, we need a trajectory that goes from \((x_0,y_0)\) to
\((x_3, y_2)\), iterating over the intermediate grid points (see figure
\ref{fig:steps}).

In the first step we go from \((x_0, y_0)\) to \((x_1, y_0)\). The new
potential is thus (using \eqref{eq:Iterator}):

\begin{equation*}
V(x_1, y_0) \approx V(x_0, y_0) + \Delta V(x_1, y_0; x_0, y_0)
\label{eq:IteratorEx1}
\end{equation*}

The next two steps continue in the horizontal direction, all the way to
\((x_3, y_0)\). The value of the potential there is:

\begin{equation*}
V(x_3, y_0) \approx V(x_0, y_0) + \Delta V(x_1, y_0; x_0, y_0) + \Delta V(x_2, y_0; x_1, y_0) + \Delta V(x_3, y_0; x_2, y_0)
\label{eq:IteratorEx2}
\end{equation*}

Now, to reach our destination \((x_3, y_2)\) we have to move two steps
in the vertical direction:

\begin{equation*}
\begin{split}
V(x_3, y_2) \approx V(x_0, y_0) + \Delta V(x_1, y_0; x_0, y_0) + \Delta V(x_2, y_0; x_1, y_0) + \Delta V(x_3, y_0; x_2, y_0) + \\
+ \Delta V(x_3, y_1; x_3, y_0) + \Delta V(x_3, y_2; x_3, y_1)
\end{split}
\label{eq:IteratorEx3}
\end{equation*}

Generalizing the previous example we see that we can compute the
approximate potential at a generic point \((x_i, y_j)\) using the closed
formula \eqref{eq:NumericalRecipe}. Both our example \eqref{eq:IteratorEx3}
and formula \eqref{eq:NumericalRecipe} have been derived sweeping first in
the horizontal direction and next in the vertical one. Of course, we can
choose different paths of summation. Nevertheless, because we are
building our potential neglecting the non-gradient part of our vector
field, we know that our results will converge to the same solution
regardless of the chosen path.

\section*{References}\label{references}
\addcontentsline{toc}{section}{References}

\hypertarget{refs}{}
\hypertarget{ref-Beisner2003}{}
Beisner, BE, DT Haydon, and K. Cuddington. 2003. ``Alternative stable
states in ecology.'' \emph{Frontiers in Ecology and the Environment} 1
(7). John Wiley \& Sons, Ltd: 376--82.
doi:\href{https://doi.org/10.1890/1540-9295(2003)001\%5B0376:ASSIE\%5D2.0.CO;2}{10.1890/1540-9295(2003)001{[}0376:ASSIE{]}2.0.CO;2}.

\hypertarget{ref-Bhattacharya2011}{}
Bhattacharya, Sudin, Qiang Zhang, and Melvin E Andersen. 2011. ``A
deterministic map of Waddington's epigenetic landscape for cell fate
specification.'' \emph{BMC Systems Biology} 5 (1): 85.
doi:\href{https://doi.org/10.1186/1752-0509-5-85}{10.1186/1752-0509-5-85}.

\hypertarget{ref-Edelstein-Keshet}{}
Edelstein-Keshet, Leah. 1988. \emph{Mathematical Models in Biology}.
Society for Industrial; Applied Mathematics.
doi:\href{https://doi.org/10.1137/1.9780898719147}{10.1137/1.9780898719147}.

\hypertarget{ref-escher}{}
Escher, Maurits Cornelis. 1960. ``Klimmen en dalen.''

\hypertarget{ref-escher1961}{}
---------. 1961. ``Waterval.''

\hypertarget{ref-Freidlin2012b}{}
Freidlin, Mark I., and Alexander D. Wentzell. 2012. \emph{Random
Perturbations of Dynamical Systems}. Vol. 260. Grundlehren Der
Mathematischen Wissenschaften. Berlin, Heidelberg: Springer Berlin
Heidelberg.
doi:\href{https://doi.org/10.1007/978-3-642-25847-3}{10.1007/978-3-642-25847-3}.

\hypertarget{ref-Gilbert1991}{}
Gilbert, Scott F. 1991. ``Epigenetic landscaping: Waddington's use of
cell fate bifurcation diagrams.'' \emph{Biology \& Philosophy} 6 (2).
Kluwer Academic Publishers: 135--54.
doi:\href{https://doi.org/10.1007/BF02426835}{10.1007/BF02426835}.

\hypertarget{ref-Hanggi1990}{}
Hänggi, Peter, Peter Talkner, and Michal Borkovec. 1990. ``Reaction-rate
theory: fifty years after Kramers.'' \emph{Reviews of Modern Physics} 62
(2). American Physical Society: 251--341.
doi:\href{https://doi.org/10.1103/RevModPhys.62.251}{10.1103/RevModPhys.62.251}.

\hypertarget{ref-VonHelmholtz1858}{}
Helmholtz, Hermann von. 1858. ``Über Integrale der hydrodynamischen
Gleichungen, welcher der Wirbelbewegungen entsprechen.'' \emph{Journal
Für Die Reine Und Angewandte Mathematik} 55: 25--55.

\hypertarget{ref-Lagrange1777}{}
Lagrange, Joseph-Louis. 1777. ``Remarques générales sur le mouvement de
plusieurs corps qui s'attirent mutuellement enraison inverse des carrés
des distances.'' Académie royale des Sciences et Belles-Lettres de
Berlin.
\href{https://archive.org/stream/oeuvresdelagrang04lagr\%7B/\#\%7Dpage/396/mode/2up}{https://archive.org/stream/oeuvresdelagrang04lagr\{\textbackslash{}\#\}page/396/mode/2up}.

\hypertarget{ref-Marsden2003}{}
Marsden, Jerrold E., and Anthony Tromba. 2003. \emph{Vector calculus}.
W.H. Freeman.

\hypertarget{ref-Nolting2015}{}
Nolting, Ben Carse, and Karen C Abbott. 2015. ``Balls, cups, and
quasi-potentials: quantifying stability in stochastic systems.''
\emph{Ecology}, December, 15--1047.1.
doi:\href{https://doi.org/10.1890/15-1047.1}{10.1890/15-1047.1}.

\hypertarget{ref-Pawlowski2006}{}
Pawlowski, Christopher W. 2006. ``Dynamic Landscapes, Stability and
Ecological Modeling.'' \emph{Acta Biotheoretica} 54 (1). Kluwer Academic
Publishers: 43--53.
doi:\href{https://doi.org/10.1007/s10441-006-6802-6}{10.1007/s10441-006-6802-6}.

\hypertarget{ref-Penrose1958}{}
Penrose, L. S., and R. Penrose. 1958. ``Impossible objects: A special
type of visual illusion.'' \emph{British Journal of Psychology} 49 (1).
John Wiley \& Sons, Ltd (10.1111): 31--33.
doi:\href{https://doi.org/10.1111/j.2044-8295.1958.tb00634.x}{10.1111/j.2044-8295.1958.tb00634.x}.

\hypertarget{ref-Rodriguez-Sanchez2019}{}
Rodríguez-Sánchez, Pablo. 2019. ``PabRod/rolldown: 1.0.0,'' March.
doi:\href{https://doi.org/10.5281/ZENODO.2591551}{10.5281/ZENODO.2591551}.

\hypertarget{ref-Scheffer2001}{}
Scheffer, Marten, Steve Carpenter, Jonathan a Foley, Carl Folke, and
Brian Walker. 2001. ``Catastrophic shifts in ecosystems.'' \emph{Nature}
413 (6856): 591--96.
doi:\href{https://doi.org/10.1038/35098000}{10.1038/35098000}.

\hypertarget{ref-Strogatz1994}{}
Strogatz, Steven H. 1994. \emph{Nonlinear Dynamics and Chaos: With
Applications to Physics, Biology, Chemistry and Engineering}.

\hypertarget{ref-Volterra1926}{}
Volterra, Vito. 1926. ``Variazioni e fluttuazioni del numero d'individui
in specie animali conviventi.'' \emph{Memorie Della R. Accademia Dei
Lincei} 6 (2): 31--113.
doi:\href{https://doi.org/10.1093/icesjms/3.1.3}{10.1093/icesjms/3.1.3}.

\hypertarget{ref-Zhou}{}
Zhou, Joseph Xu, M. D. S. Aliyu, Erik Aurell, and Sui Huang. 2012.
``Quasi-potential landscape in complex multi-stable systems.''
\emph{Journal of the Royal Society Interface} 9 (77): 3539--53.
doi:\href{https://doi.org/10.1098/rsif.2012.0434}{10.1098/rsif.2012.0434}.

\end{document}